\documentclass[aps, prd, amsmath, floats,floatfix, twocolumn, superscriptaddress, nofootinbib, showpacs]{revtex4}

\usepackage{amssymb}
\usepackage{amsmath}
\usepackage{verbatim}
\usepackage{mathrsfs}
\usepackage{amsfonts}
\usepackage{latexsym}
\usepackage{epsfig}
\usepackage{color}
\usepackage{graphicx,subfigure}
\usepackage{units}
\usepackage{slashbox}

\newcommand{\be}{\begin{equation}}
\newcommand{\ee}{\end{equation}}

\begin{document}

\title{Cosmological production of ultralight dark matter axions}

\author{Alberto Diez-Tejedor}
\affiliation{Departamento de F\'isica, Divisi\'on de Ciencias e Ingenier\'ias,
Campus Le\'on, Universidad de Guanajuato, Le\'on, 37150, M\'exico}

\author{David J. E. Marsh}
\affiliation{Department of Physics, King's College London, Strand, London, WC2R 2LS, UK}

\begin{abstract} 

The highly populated, low-energy excitations of a scalar field of mass $m_a\sim 10^{-22}\,\textrm{eV}$ can represent the full dark matter content of the 
universe and alleviate some tensions in the standard cosmological scenario on small-scales. This {\it fuzzy dark matter} component is commonly assumed to
arise as the consequence of a new axion-like particle in the matter sector, yet for simplicity it is usually modeled in terms of a simple free quadratic field.
In this paper we consider how the cosmological constraints are modified when the effects of an instanton potential and temperature-dependent mass are included. 
Current isocurvature and tensor bounds confirms that this particle should be formed before the end of a low-scale inflation period with Hubble parameter 
$H_I\lesssim 2.5\times 10^{12}\,\textrm{GeV}$, in accordance with previous free-field analysis. The axion decay constant, $f_a$, which fixes axion couplings, 
appears in the instanton potential and determines the relic abundance, the stability of galaxy cores to axion emission, and the direct searches of fuzzy dark matter. 
If the axion mass is $T$-independent, we find that $f_a\gtrsim 10^{16}\,\text{GeV}$ is required in order to reach the observed relic density without fine tuning the 
initial conditions, while for a $T$-dependent case this bound can be lowered by an order of magnitude. However, the anharmonicities in the instanton potential, and mainly a 
$T$-dependent mass, can delay the onset of field oscillations, leading to larger physical suppression scales in the matter power spectrum for a fixed zero-temperature 
axion mass. This may favor a string or accidental axion over one emerging from a strongly coupled gauge sector if this model is required to provide large galactic halo 
cores while simultaneously satisfying observational constraints from cosmic structure formation.

\end{abstract}


\maketitle

\section{Introduction}

New numerical cosmological simulations support the idea of a scalar particle of mass $m_a\sim 10^{-22}\,\textrm{eV}$ 
representing the full dark matter (DM) content of the universe~\cite{simulations} ---see Refs.~\cite{BECDM} for related work, and~\cite{review, Hui} for 
recent reviews. At large scales the coherent macroscopic excitation of an ultralight scalar particle can mimic the behavior of a cold dark matter 
(CDM) component~\cite{misalignment}. At the scale of galaxies, however, this new particle could alleviate by means of its macroscopic wave-like properties 
some of the classic ``discrepancies'' of the standard cosmological scenario ~\cite{review, Hui, Hlozek15, Hlozek16, Alma2}.

In this paper we explore the cosmological production of ultralight scalar DM particles with a mass of the order $m_a\sim 10^{-22}\,\textrm{eV}$.
We concentrate to the case in which all the particles are generated through a vacuum misalignment in the early universe~\cite{misalignment}.
Other contributions could exist coming from the decay of cosmic strings and/or thermal relics, see e.g. Refs.~\cite{string_decay, thermal_axions}. However, as we will 
find next, in order to be consistent with the non-observation of tensor modes in the cosmic microwave background (CMB) it is necessary that such contributions 
be diluted to almost zero during inflation, and for that reason we do not need to consider them here.

It has long been known that light scalar particles bring (un)naturalness issues. These issues could be alleviated, however, if we think 
on those fields as pseudo Nambu-Goldstone (pNG) bosons. The QCD axion, introduced by Peccei and Quinn~\cite{PQ} to solve the strong CP problem in quantum chromodynamics (QCD), 
represents a particular realization of a pNG scenario~\cite{KolbTurner, Sikivie}, but a plethora of light scalar particles appear also in the context of 
the {\it string axiverse}~\cite{axiverse, pNG}. An axion-like particle is characterized by two energy scales. On the one hand the axion decay constant, $f_a$, 
which may be also related to the scale at which a global Peccei-Quinn (PQ)-like symmetry is broken spontaneously. On the other, the scale at which the symmetry is broken explicitly, $\mu$,
generically due to the presence of global anomalies in the quantum theory. These two energy scales contribute to the mass of the scalar particle in the form $m_a\sim \mu^2/f_a$~\cite{Wilczek1978}.

In terms of galaxy properties, the most important factor is simply this mass coefficient. Current astrophysical analysis are becoming increasingly sophisticated and place interesting bounds on the 
particle mass~\cite{simulations, review, Hui, Alma}, but they are otherwise insensitive to the fine details of the model.
In this paper we make a connection between particle physics model parameters, the axion relic abundance, structure formation, and the theory of inflation to see if these considerations can drastically alter 
the interpretation of the previous astrophysical modeling. 

Three parameters mainly determine the (non-thermal) relic abundance of axion particles~\cite{Gondolo1, Gondolo2, Wilczek, Wantz, Marshbicep}: 
the axion decay constant, $f_a$, the Hubble parameter at the end of inflation, $H_I$, and the initial value of the misalignment angle, $\theta_i$. Here 
we constrain the model parameters assuming that the ultralight particles make up the totality of the CDM in the universe, 
using the the relic density and the upper bounds on the uncorrelated isocurvature and tensor amplitudes as reported by \emph{Planck} (2015) in Ref.~\cite{Planck}
(a full study using \textsc{axionCAMB}~\cite{Hlozek15,Hlozek16} is beyond the scope of this work).

In general the relic abundance of an ultralight scalar is sensitive to the details of how the mass term is generated at the scale of the explicit symmetry breaking. 
We parameterize this process via the temperature dependence of the axion mass, and show that general conclusions can still 
be drawn despite unknown specifics of the model. All that we assume is a period of cosmic inflation, but otherwise the conclusions are not sensitive to the details of 
what happened before Big-Bang nucleosynthesis (BBN).

Previous arguments based on an idealized misalignment production point to an axion decay constant of order $f_a\sim 10^{17}\,\text{GeV}$, but if this parameter 
could be consistently lowered by the nonlinear/temperature-dependent properties of the potential, this may have important consequences both for astrophysics~\cite{Levkov, Helfer} 
and direct detection probes~\cite{casper}. Another interesting consequence of an instanton potential and a $T$-dependent mass is that, in general, the 
characteristic suppression scale on the mass power spectrum is not necessarily correlated with the size of galaxy cores in this model. This is because the matter density profiles of 
galaxies is essentially only sensitive to the zero-temperature mass of the particle, whereas the nonlinear and $T$-dependent properties of the potential can 
move the position of the cut-off in the power spectrum to smaller wavenumbers. These and other interesting cosmological signatures of a realistic axion potential will be 
discussed along the text.

The paper is organized as follows. In Section~\ref{sec.model} we overview the model and introduce our mass parametrization. We use this to codify 
possible observational signatures coming from a realistic potential term, at the level of the cosmological background, Section~\ref{sec.background}, as well as the linear 
perturbations, Section~\ref{sec:structure}. Then in Section~\ref{sec.constraints} we use the observed value of the relic density, the current upper bounds for the amplitudes 
of tensor and isocurvature modes, and some general arguments on structure formation to constrain the parameters of the model.
Finally we conclude in Section~\ref{sec.discussion} with a discussion. Some additional information regarding how the cosmological observables are related to a period of 
inflation, and the effects of the nonlinear terms on the axion mass density can be found in the appendices.

\section{The model}\label{sec.model}

The  mass of a scalar particle $m$ gets radiative contributions from its interactions with the other constituents of 
the standard model of particle physics, and grows quadratically with the highest energy scale $\Lambda_{\textrm{SM}}$ at which 
the model is valid, $m^2\to m^2+\mathcal{O}(1)\Lambda_{\textrm{SM}}^2$~\cite{Peskin}. It is then difficult to imagine a  
scalar particle of mass $m_s\ll \Lambda_{\textrm{SM}}$, where $\Lambda_{\textrm{SM}}$ is usually identified with the Planck scale.
Symmetries can alleviate this situation. For instance, supersymmetry transforms the quadratic running of the mass to a logarithmic one, 
making it easier to understand e.g. why the mass of the Higgs boson is so low when compared to the Planck scale~\cite{feng}.
However, even in this scenario, the masses of the scalar particles are expected to be of the order of the supersymmetry-breaking-scale, 
$\Lambda_{\textrm{SUSY}}\sim 1\, \textrm{TeV}$, very far from the $10^{-22}\,\textrm{eV}$ considered in Refs.~\cite{simulations}.

Scalar particles of masses much lower than this scale can naturally appear, however, in the context of the pNG bosons~\cite{PQ}. In its simplest realization a global $U(1)$ 
symmetry is spontaneously broken at an energy scale $f_a$ by the vacuum expectation value of a complex scalar, $\langle\phi(x)\rangle=(f_a/\sqrt{2}) e^{i\theta(x)}$.
The effective theory describing the Goldstone bosons is invariant under shift transformations, $\theta\to \theta+\textrm{const}$, and then these particles are massless.
(The excitations of the radial mode have mass of order $f_a$ and do not contribute to the low energy spectrum.)
Furthermore, if the whole underlying theory respects the $U(1)$ symmetry, then all the couplings of the Goldstones to additional fields also needs to satisfy this shift symmetry, 
and thus perturbative quantum effects cannot give mass to these particles.

Instanton effects and other non-perturbative physics, however, can break this symmetry explicitly down to $\theta\to \theta+2\pi/N$ (for some integer $N$), leading a potential term 
$\mu^4V(\theta)$ to the field~$\theta$. With no loss of generality one usually choses the minimum of the potential at $\theta=0$, with $-\pi/N<\theta<\pi/N$ measuring 
the misalignment angle with respect to the equilibrium state. Here $\mu$ is the scale of the explicit symmetry breaking, and by construction $V(\theta)=V(\theta+2\pi/N)$. 
The potential provides the would-be-Goldstone particles of an effective mass, that is now protected against radiative corrections from any possible interaction with the other particles.
If we canonically normalize the kinetic term, and absorbing the integer $N$ into the axion decay constant and misalignment angle (i.e. from now on we will use $f_a$ and $\theta$
to denote $f_a/N$ and $N\theta$, respectively), the effective low energy degrees of freedom in the theory can be finally described in terms of the Lagrangian density
\begin{equation}
 \mathcal{L} =\frac{1}{2} \partial_{\mu}a\partial^{\mu}a-\mu^4\left[1-\cos\left(\frac{a}{f_a}\right)\right], 
\end{equation}
where $a=f_a\theta$ is the axion-like pNG particle. Expanding the potential term to the quadratic order in the axion field we can easily identify $m_a= \mu^2/f_a$
as the mass of the pseudoscalar particle. Note that the particular cosine form of the potential holds strictly in the semiclassical or dilute gas limits
at lowest order in the instanton expansion, and for the QCD axion a more accurate potential can be obtained in chiral perturbation theory~\cite{Veneziano, Hardy}. 

In the case of the QCD axion we have that $\mu^4\approx m_u\Lambda_{\rm QCD}^3$. Here $m_u$ is the u-quark mass and $\Lambda_{\rm QCD}$ the QCD strong-coupling scale, and then 
$\mu\approx 200\,\textrm{MeV}$. The axion decay constant $f_a$, however, is only mildly constrained by standard model interactions, leading to a range of allowed masses 
$10^{-10}\,\text{eV}\lesssim m_a\lesssim 10^{-3}\,\text{eV}$~\cite{Raffelt,Huang}. For a generic axion the parameters $\mu$ and $f_a$ are both in principle arbitrary, but they are related 
by the actual mass of the scalar particle, expected to be of the order $\mu^2/f_a\sim 10^{-22}\,\textrm{eV}$~\cite{simulations}. 
If one imposes the condition $f_a\lesssim m_{\textrm{Pl}}\approx 2.431\times 10^{18}\,\textrm{GeV}$ (where $m_{\textrm{Pl}}=M_{\textrm{Pl}}/\sqrt{8\pi}$ is the reduced Planck mass), 
this restricts the scale of the explicit symmetry breaking to be less than around a~$\textrm{keV}$ to obtain the desired mass. 

This scenario where non-perturbative physics is behind the appearance of a scalar particle of light mass provides one possible realization of the general picture we want to explore 
in this paper. However, there are at least two other possibilities for generating an ultralight axion-like particle, both giving rise to an effectively $T$-independent potential. The first one 
(also in the context of a field theory) follows the template of the {\it accidental} axion, where the global $U(1)$ symmetry is not exact but, instead, arises at leading order 
as a consequence of an underlying discrete $Z_N$ symmetry, e.g.~\cite{KimMarsh, Dias}. The potential is now generated at some UV energy scale $\Lambda_{\textrm{UV}}$ that breaks the 
global symmetry, however, since the symmetry breaking operators appear only at very high order, we expect that $\mu\ll \Lambda_{\rm UV}$, lower orders being forbidden by the 
discrete symmetry. Again this guarantees the appearance of light scalar particles, but in this case the potential is independent of the temperature in the low-energy 
theory at temperatures of order $T\sim \mu$. 

The second possibility for an effectively $T$-independent potential arises in string/M-theory and other higher dimensional supergravity theories~\cite{axiverse}. In such models 
the axion fields descend from the geometry as the pseudoscalar partners of the moduli $\sigma$ demanded by supersymmetry. The moduli acquire potentials at the 
supersymmetry-breaking-scale, $\Lambda_{\rm SUSY}\sim m_{3/2}\gtrsim 1\text{ TeV}$, where $m_{3/2}$ is the gravitino mass. The axion, however, acquires a mass term much below this 
scale, with $\mu\sim \sqrt{m_{3/2}M_{\textrm{Pl}}}e^{-\mathcal{Q}\sigma}\ll m_{3/2}$. Here $\mathcal{Q}\sim 2\pi/\alpha_{\rm GUT}$ is an instanton charge, and $\alpha_{\rm GUT}\sim 1/25$
the grand unified coupling~\cite{Hui, Acharya}, with $\sigma\sim \mathcal{O}(1)$. 
The instanton potential may be resummed assuming modular invariance as a Dedekind eta function, e.g. Ref.~\cite{Kobayashi},
with any temperature dependence restricted to the supersymmetry breaking dynamics in the UV. 

Each of the three scenarios above provide possible particular realizations of the DM axion discussed in Refs.~\cite{simulations}. The temperature dependence of the potential can affect the 
axion relic abundance, and we must account for this possibility. For instance, consider the case of a potential generated by the transition to a strong coupling regime, 
following the template of the QCD axion. Then, at temperatures much lower than the energy scale of the explicit symmetry breaking, $T\ll \mu$,
we know that $m_a(T)=m_a(T=0)=m_a$. At higher temperatures, $\mu\ll T\ll f_a$, on the contrary, the axion is effectively massless, $m_a(T)= 0$. 
The task is then to model how these two asymptotic regimes are connected at temperatures of the order $T\sim \mu$. 

\begin{figure}
 \includegraphics[width=0.49\textwidth]{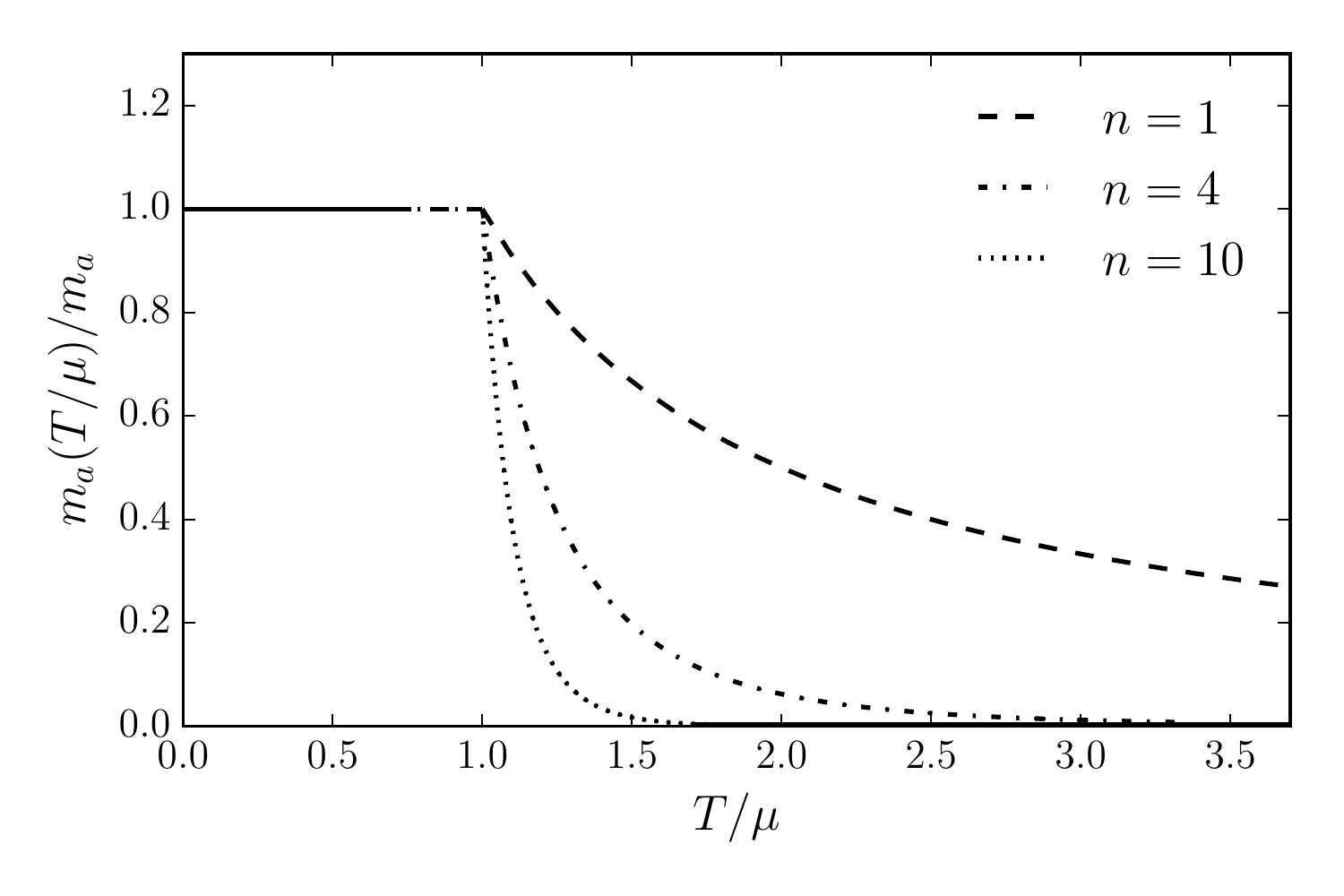}
 \caption{{\bf Mass Parameter Vs Temperature.} Parametrization of the mass term as a function of the temperature, Eq.~(\ref{eq:mass}), for different values of the constant $n$. 
 Note that $m_a(T\ll \mu)=m_a$, and $m_a(T\gg \mu)=0$, solid lines in the figure. The parameter $n$ measures the (unknown) sharpens of the transition, 
 represented here in terms of discontinuous lines.}
 \label{figure:mass}
\end{figure}

To proceed we will make the following choice which parameterizes the range of possibilities for the origin of the scalar potential:
\begin{equation}
  \label{eq:mass}
  m_a(T)= \left\lbrace 
    \begin{array}{ll}
      m_a\left(\frac{\mu}{T}\right)^n \; & \textrm{for} \;
      T> \mu,  \\
      m_a  \; & \textrm{for} \; T\leq \mu. 
    \end{array}
  \right. 
\end{equation}
The coefficient $n$ in Eq.~\eqref{eq:mass} codifies the sharpness of the phase transition: for $n=0$ the mass term is independent of the temperature, 
whereas for $n\gg1$ the change of $m_a(T)$ is abrupt at $T\sim \mu$; see Figure~\ref{figure:mass} for details. The value of the axion relic density depends on the parameter $n$. 
However, as we will show next, this quantity is not very sensitive to the peculiarities of the potential at temperatures around the scale of the explicit symmetry breaking, 
although 
these fine details could have leaved some imprints on large scale structure accessible to observations.

The value of the index $n$ can be computed once a specific particle content for the sector giving rise to the axion mass is fixed: gauge group of the strongly coupled sector, 
number of fermions, etc. For the QCD axion, the most simple ``dilute instanton gas'' calculation gives $n=4$~\cite{dig}, while the ``interacting instanton liquid'' 
$n\approx 3.34$~\cite{Wantz}, and lattice calculations $n=3.55\pm 0.30$~\cite{lattice}. For a string or an accidental axion, however, $n=0$.

While the present work was in preparation, Ref.~\cite{ULA_newmodel} appeared, presenting a model for ultralight axions in a hidden-sector copy of QCD dubbed $\mu$-QCD, 
with index $n=11/3$ in the dilute instanton gas approximation under the assumption of a minimal matter content. Additional sterile neutrinos and a period of mild thermal inflation are necessary to avoid dangerous relics, 
and the model predicts $\Delta N_{\rm eff}\approx 0.25$ extra relativistic degrees of freedom in the CMB, which could easily be detected by CMB-S4~\cite{s4scibook}.

\section{Cosmology}\label{cosmology}

An instanton potential and a $T$-dependent mass can modify previous constraints on the model~\cite{review}.
Now we use the parametrization in Section~\ref{sec.model} to explore how the new terms in the potential can affect ultralight axion cosmology.

\subsection{Background evolution}\label{sec.background}

In a homogeneous and isotropic universe the dynamical evolution of the misalignment angle is described in terms of the equation
\begin{equation}
\label{eq:KG}
 \ddot{\theta}+3H(T)\dot{\theta}+m^2_a\theta=0.
\end{equation}
Here the dot denotes the derivative with respect to the cosmic time, $H(T)$ is the Hubble parameter as a function of the temperature, and $m_a$ is the 
mass of the scalar particle. Note that the expression in Eq.~(\ref{eq:KG}) is valid only for small values of the misalignment angle, $|\theta|\lesssim 1$, when 
the potential is approximately harmonic, and as long as the axion mass does not vary with the temperature. The nonlinear $T$-dependent contributions will be 
considered later in this section and in more detail in Appendix~\ref{appendix2}.

There are two different regimes for the solutions to Eq.~(\ref{eq:KG}), depending on the relative values of the Hubble expansion rate, $H(T)$, and 
mass parameter, $m_a$. At high temperatures the scalar particle is effectively massless. There are two linearly independent solutions
to Eq.~(\ref{eq:KG}) when $m_a=0$: one of them decays with the cosmological expansion and is irrelevant for our considerations
in this paper; the other remains constant and fixes the ``initial'' value of the misalignment angle, $\theta=\theta_i$.

In the linear theory the axion field is frozen at $\theta=\theta_i$ until the mass and Hubble terms satisfy
\begin{equation}
\label{eq:start.oscillations}
 3H(T_{\textrm{osc}}^{\textrm{free}})=m_a
\end{equation}
and the misalignment angle starts to oscillate. Here as in many other places along the paper the equal sign should be taken as an indicative of an ``order of magnitude.''
For a DM component this necessarily occurs during radiation domination, so we can approximate 
\begin{equation}
\label{eq:H(T)}
 H(T)=\left[\frac{8\pi^3g_{*}(T)}{90M_{\textrm{Pl}}^2}\right]^{1/2}T^2= 1.66g^{1/2}_{*}(T)\frac{T^2}{M_{\textrm{Pl}}},
\end{equation}
where $g_{*}(T)$ is the total number of effectively massless degrees of freedom contributing to the energy density~\cite{KolbTurner}. 
As we argue next, we can concentrate to temperatures below a MeV, and then write $g_{*}(T)\approx 3.36$.  
Introducing the expression for the Hubble parameter in Eq.~(\ref{eq:H(T)}) into Eq.~(\ref{eq:start.oscillations}), the temperature $T_{\textrm{osc}}^{\textrm{free}}$ 
gets fixed to
\begin{equation}
  T_{\textrm{osc}}^{\textrm{free}}= 366m_{22}^{1/2}\,\textrm{eV}.\label{eq:T1}
\end{equation}
Here and below we use 
$m_a=m_{22}\times 10^{-22}\,\textrm{eV}$.

An instanton potential and a $T$-dependent mass can make that the temperature $T_{\textrm{osc}}^{\textrm{free}}$ 
at which $3H(T_{\textrm{osc}}^{\textrm{free}})=m_a$ does not necessarily coincide with the temperature for the onset of field oscillations, $T_{\textrm{osc}}$, and there could be 
some delay in $T_{\textrm{osc}}$ with respect to the value in Eq.~(\ref{eq:T1}) if the initial misalignment angle is large enough, $|\theta_i|\gtrsim 1$, or if the axion mass varies 
significantly with the temperature around $T\sim T_{\textrm{osc}}$, $n\neq 0$ in Eq.~(\ref{eq:mass}). At the effective level it is possible to parametrize these nonlinear $T$-dependent effects in terms of an 
expression of the form 
\begin{equation}\label{def.Tosc}
 T_{\textrm{osc}}= F_T(\theta_i, f_a, n)T^{\textrm{free}}_{\textrm{osc}},
\end{equation}
where
\begin{widetext}
\begin{equation}\label{eq.fT}
 F_T(\theta_i,f_a,n)= \left[f(\theta_i)\right]^{-1/2}\left\lbrace \begin{array}{ll}
      1, \; & \textrm{for }
      f_a \ge f_a^*, \textrm{ or } f_a<f_a^* \textrm{ and } \theta_i\ge \theta_i^*, \\
      \left[(f_a/f_a^*)f(\theta_i)\right]^{\frac{n}{2(n+2)}},  \; & \textrm{for } f_a < f_a^* \textrm{ and } \theta_i< \theta_i^*.
    \end{array} 
    \right. 
\end{equation}
\end{widetext}
Here $f_a^*= 0.550\,m_{\textrm{Pl}}$ is a characteristic scale in the theory for the axion decay constant, and $\theta_i^*$ a distinctive value for the initial misalignment  
angle to be defined next in Eq.~(\ref{eq.angle}). Different authors have considered different expressions for the function $f(\theta_i)$, see for 
instance~\cite{Turner1986,Lyth, Weiler, Bae}. However, the qualitative features of the final results are not very sensitive to the particular choice. 
We find it convenient to work in terms of 
\begin{equation}\label{eq.def.f}
 f(\theta_i)=\ln\left[\frac{e}{1-(\theta_i/\pi)^4}\right],
\end{equation}
which sets the characteristic angle $\theta^*_i$ to
\begin{equation}\label{eq.angle}
 \theta_i^* = \pi \left[1-\exp\left(1-f_a^*/f_a\right)\right]^{1/4}.
\end{equation}
With no loss of generality we have assumed $0<\theta_i<\pi$. More details are given in Appendix~\ref{appendix2}.

Note that if the initial misalignment angle is small, $\theta_i\ll \pi$, then $f(\theta_i)\approx 1$, and the effect of the nonlinear terms in the potential is negligible, 
as expected. If on the contrary the initial angle approaches the critical value at $\theta_i=\pi$, the function $f(\theta_i)$ in Eq.~(\ref{eq.def.f}) grows indefinitely, and the delay 
due to the anharmonicities can be very large. 

Another possible source of delay in $T_{\textrm{osc}}$ comes from a $T$-dependent mass in the potential term. 
If $f_a\ge f_a^*$, i.e. $\mu\ge 366m_{22}^{1/2}\,\textrm{eV}$, the shift symmetry is broken so early in the history of the universe that 
the value of $T_{\textrm{osc}}$ is not sensitive to the temperature dependence of the potential, and $T_{\textrm{osc}}=[f(\theta_i)]^{-1/2}T_{\textrm{osc}}^{\textrm{free}}$. 

However, for $f_a<f_a^*$, i.e. $\mu< 366m_{22}^{1/2}\,\textrm{eV}$, the value of $T_{\textrm{osc}}$ depends crucially on the particular choice of $m_a(T)$ and the initial value of the
misalignment angle. In terms of the parametrization in Eq.~(\ref{eq:mass}), we obtain that, if $n=0$, the mass term does not change with the temperature, and again any variation of 
$T_{\textrm{osc}}$ with respect to $T_{\textrm{osc}}^{\textrm{free}}$ should come at the expense of the anharmonicities in the potential, $T_{\textrm{osc}}=[f(\theta_i)]^{-1/2}T_{\textrm{osc}}^{\textrm{free}}$. 
If on the contrary $n\gg1$, the change in the mass is so sharply located around the scale of the explicit symmetry breaking that, as long as the initial misalignment angle 
remains relatively small, $\theta_i<\theta_i^*$, it is the scale of the breakdown of the shift symmetry that determines the onset of field oscillations, $T_{\textrm{osc}}\approx \mu$.
If the initial angle is large enough, $\theta_i\ge\theta_i^*$, however, the onset of field oscillations cannot feel the variations in the mass, even when $f_a<f_a^*$, and 
$T_{\textrm{osc}}=[f(\theta_i)]^{-1/2}T_{\textrm{osc}}^{\textrm{free}}$.
Note that unless $f_a\sim f_a^*$, we can approximate $\theta_i^*=\pi$ in Eq.~(\ref{eq.angle}), so this last possibility is very unlikely in practice for low $f_a$ models.
We can then conclude that, unless the parameter $f_a$ lies close to the Planck scale, the temperature dependence of the mass is always relevant.
Furthermore, e.g. $f(\theta_i=3.14)=7.20$, not very far from unity, so the $T$-dependent terms usually dominate unless there were an extreme fine tuning of the initial conditions.

The expression in Eq.~(\ref{def.Tosc}) is valid as long as the mass of the scalar particle is less than about $10^{-15}\,\textrm{eV}$, and then it is general for ultralight axion
DM candidates, but it does not apply for e.g. the QCD axion. For particles with a mass larger than $10^{-15}\,\textrm{eV}$ the value of $T_{\textrm{osc}}$ grows above a MeV, and 
we need to take into account that the effective number of relativistic species $g_*(T)$ varies with the temperature. For lighter candidates, however, the temperature at the onset of 
field oscillations is always less than a few hundred of $\textrm{eV}$, so the axion is still frozen at BBN, $T_{\textrm{BBN}}\sim 1\,\textrm{MeV}$. This guarantees the validity 
of the expression in Eq.~(\ref{eq:H(T)}), with $g_{*}(T)\approx 3.36$, and then the results of this paper are not sensitive to the unknown details of what happened 
between inflation and BBN, up to uncertainties on $\Delta N_{\rm eff}$, e.g. ~\cite{Acharya}. 

Let us compute the contribution of the misaligned vacuum to the matter content of the universe as measured by an observer comoving with the expansion. In order to do so we need 
to obtain first the number density of particles around $T_{\textrm{osc}}$ using the expression~\cite{Gondolo1, Gondolo2, Wilczek, Wantz, Marshbicep, Turner1986, Lyth}
\begin{equation}
\label{eq:n.a}
 n_a^{\textrm{mis}}(T_{\textrm{osc}})=\frac{1}{2}m_a(T_{\textrm{osc}})f_a^2\langle\theta_i^2\rangle ,
\end{equation}
where the brackets makes reference to some spatial averages to be defined soon. From Eq.~(\ref{eq:n.a}), and imposing the conservation of the comoving axion number density with 
the cosmological expansion once field oscillations start, $n_a^{\textrm{mis}}(T)/s(T)=\textrm{const}$, we obtain that the energy density in axions today, $\rho_a^{\textrm{mis}}(T_0)=m_a(T_0) n_a^{\textrm{mis}}(T_0)$, 
is determined by the expression
\begin{equation}
\label{eq:densiti.today}
 \rho_a^{\textrm{mis}}(T_0)=\frac{1}{2}\frac{T_0^3}{T_{\textrm{osc}}^3}m_a m_a(T_{\textrm{osc}})f_a^2\langle\theta_i^2\rangle ,
\end{equation}
with $m_a(T_0)=m_a$. Here $T_0\approx 2.7255\,\textrm{K}$ is the temperature of the CMB photons at the present time~\cite{fixsen}, and we have used the expression 
$s(T)=(2\pi^2/45)g_{*S}(T)T^3$ for the entropy density in the universe, where $g_{*S}(T)$ counts for the total number of effectively massless degrees of freedom contributing to the 
entropy density. In the standard cosmological scenario and for the low temperatures of interest this quantity remains constant and fixed to $g_{*S}(T)\approx 3.91$~\cite{KolbTurner}. 
Dividing Eq.~(\ref{eq:densiti.today}) by the the critical density today, $\rho_c=3H_0^2/(8\pi G)$, and taking into account the expression in Eq.~(\ref{def.Tosc}) for $T_{\textrm{osc}}$,  
we finally obtain
\begin{equation}
  \label{eq:Omega}
  \Omega_a^{\textrm{mis}}h^2= 29.3\, m_{22}^{1/2}(f_a/f_a^*)^2  \langle F_{n_a}(\theta_i,f_a,n)\theta_i^2\rangle .
\end{equation}
Here $\Omega_a^{\textrm{mis}}\equiv\rho_a^{\textrm{mis}}(T_0)/\rho_c$ is the dimensionless density parameter and, as usual, we have parametrized the Hubble constant today in the 
form $H_0=100h\, \textrm{(km/s)/Mpc}$. The function $F_{n_a}(\theta_i, f_a,n)$ is defined from Eq.~(\ref{eq.fT}) by
\begin{equation}\label{eq.deffna}
F_{n_a}(\theta_i,f_a,n)=f(\theta_i)F_T^{-1}(\theta_i,f_a,n)
\end{equation}
and accounts for the nonlinear $T$-dependent contributions to the dynamical evolution, now at the level of the number/energy density. Apart from this function Eq.~(\ref{eq:Omega}) 
coincides with its harmonic $T$-independent counterpart, so we can then look at $F_{n_a}$ as a ``recalibration'' of the initial value of the misalignment angle necessary to obtain
the correct result. Note that even if it is not explicit in Eq.~(\ref{eq:Omega}), the dimensionless density parameter scales with $(f_a/f^*_a)^{(8+3n)/2(2+n)}$ if $f_a<f^*_a$, $\theta_i< \theta_i^*$.

The contribution of the misaligned vacuum to the energy density depends crucially on the value of $\langle F_{n_a}\theta_i^2\rangle$, and in particular on
when was the global symmetry spontaneously broken and 
the axion field formed~\cite{Wilczek}. Next we explore in some detail the two different possible scenarios.

\subsubsection{Scenario A: The symmetry is spontaneously broken after the end of inflation, $f_a<H_I/(2\pi)$}\label{sec:after}

As long as the axion field was formed after the end of inflation, the initial misalignment angle $\theta_i$ will take different values along the observable universe. 
If we consider that they are distributed randomly along the interval $(0,\pi)$, and average over all the possible outcomes, 
\begin{equation}
\label{eq:scenarioA}
 \langle F_{n_a}\theta_i^2 \rangle = \frac{1}{\pi}\int_{0}^{\pi}F_{n_a}(\theta_i,f_a,n) \theta_i^2 d\theta_i ,
\end{equation}
we obtain a number that depends on the particular values of the parameters $f_a$ and $n$, but that is otherwise a fixed constant in the theory.

\subsubsection{Scenario B: The symmetry is spontaneously broken before the end of inflation, $f_a>H_I/(2\pi)$}\label{sec:before}

According to the standard cosmological picture the whole observable universe comes from a single (nearly) homogeneous and isotropic patch  
at the end of inflation. Then, if the axion field was formed before the end of this period (and assuming that the reheating temperature does not restore the 
PQ symmetry after inflation; if so see Scenario~A), the misalignment angle will be described in terms of a unique initial value $\theta_i$. Superimposed to this value there are 
quantum fluctuations of variance $[H_I/(2\pi f_a)]^2$~\cite{Wilczek, Lyth}. Combining these two contributions we obtain
\begin{equation}
\label{eq:scenarioB}
 \langle F_{n_a}\theta_i^2 \rangle = F_{n_a}(\theta_i,f_a,n)\left[\theta_i^2+\left(\frac{H_I}{2\pi f_a}\right)^2\right]  .
\end{equation}
The quantity $H_I/(2\pi f_a\theta_i)$ is tightly constrained by the cosmological observations, see Appendix~\ref{appendix} for details, and this term will be disregarded soon. 
However, now the initial misalignment angle can take any value in the range $(0,\pi)$. Then, as long as $f_a>H_I/(2\pi)$, the quantity $\langle F_{n_a}\theta_i^2 \rangle$ in
Eq.~(\ref{eq:Omega}) is essentially a dimensionless constant, the natural range of which is determined by the allowed tuning of $\theta_i$.

\subsection{Structure formation}\label{sec:structure}

Axions and other light particles in the universe prevent the growth of cosmic structure and suppress the mass power spectrum below a characteristic length scale.
For appropriate model parameters, this length scale could be of astrophysical interest, making it possible to leave some signatures on the cosmological observables.

In the early epoch of the universe the Hubble parameter acts as a friction term that inhibits the potential gradients in the Klein-Gordon equation. 
The field is then frozen, and perturbations cannot grow. When field oscillations start, however, Jeans instability is released and cosmic structure
emerges, but those modes with a wavelength shorter than the size of the Hubble horizon at that time have been erased~\cite{KhlopovPark_etAl}.
Then, if the particles are generated through the misalignment of the vacuum in the early universe, the characteristic scale that fixes the cut-off in the mass power spectrum is 
determined by the size of the {\it Hubble horizon} at the time when field oscillations started (compare this with a thermal candidate, where this scale 
is determined by the {\it particle horizon} at the time when the thermal relic became nonrelativistic). 

The Hubble horizon (in physical coordinates) is of the order of the inverse of the Hubble parameter at a given time, $H^{-1}$. The field starts rolling during the radiation dominated
era, at $T_{\textrm{osc}}< \textrm{MeV}$, so the value of the Hubble horizon at the onset of field oscillations can be easily obtained from Eq.~(\ref{eq:H(T)}), 
making $T=T_{\textrm{osc}}$ and $g_*(T_{\textrm{osc}})\approx 3.36$. Thenceforth, this length scale has grown with the scale factor as the universe expands, 
$H^{-1}_{\textrm{osc}}(a)=H^{-1}(a_{\textrm{osc}})a/a_{\textrm{osc}}$, where $a_{\textrm{osc}}=a(T_{\textrm{osc}})$. 
For a universe in adiabatic expansion, $T(a)=T_0/a$, and $a/a_{\textrm{osc}}=T_{\textrm{osc}}/T$. Working in terms of wavenumbers instead of length scales, 
$k_{\textrm{cut-off}}= H_{\textrm{osc}}(a=1)$, we can write
\begin{equation}\label{eq.cutoff}
 k_{\textrm{cut-off}} = \left[\frac{8\pi^3g_*(T_{\textrm{osc}})}{90}\right]^{1/2}\frac{T_0T_{\textrm{osc}}}{M_{\textrm{Pl}}}.
\end{equation}
Introducing the expression for $T_{\textrm{osc}}$ as reported in Section~\ref{sec.background} into Eq.~(\ref{eq.cutoff}), and after some algebra, we finally obtain
\begin{equation}
  \label{}
  k_{\textrm{cut-off}} = F_T(\theta_i,f_a,n) k_{\textrm{cut-off}}^{\textrm{free}} ,
\end{equation}
where
\begin{equation}
 k_{\textrm{cut-off}}^{\textrm{free}}=3.35m_{22}^{1/2}\,\textrm{Mpc}^{-1}
\end{equation}
is the scale of the cut-off for a $T$-independent, harmonic potential. 

For a DM axion of mass $m_a\sim 10^{-22}\,\textrm{eV}$, the characteristic sign in the linear power spectrum at a Mpc scale has been previously identified (sometimes in an 
independent way) many times in the literature~\cite{review, Hui}. In galaxies, this same mass parameter fixes the characteristic size of the cores to the scale of a kpc. However, an instanton potential with 
a $T$-dependent mass makes it possible to release this classic correlation between cosmological and astrophysical observations. While the size of the galactic cores is only 
sensitive to the zero-temperature mass of the particle, the suppression scale on the mass power spectrum can be moved to larger physical length scales by means of the anharmonic, 
and mainly temperature-dependent, corrections to the axion potential in the early universe.
We explore this in more detail in Section~\ref{sec.constraints}.

\begin{figure*}
 \includegraphics[width=0.495\textwidth]{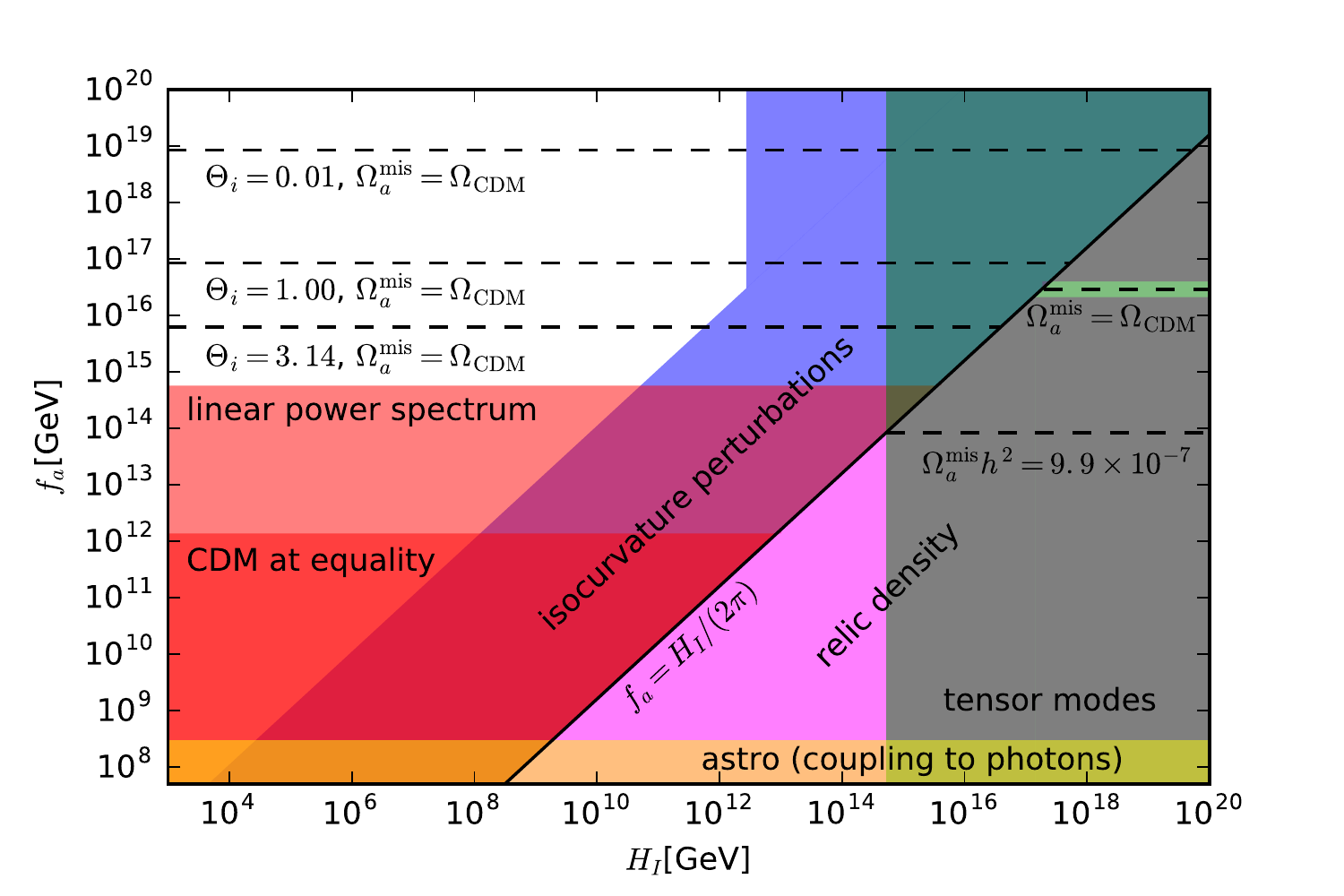}  
 \includegraphics[width=0.495\textwidth]{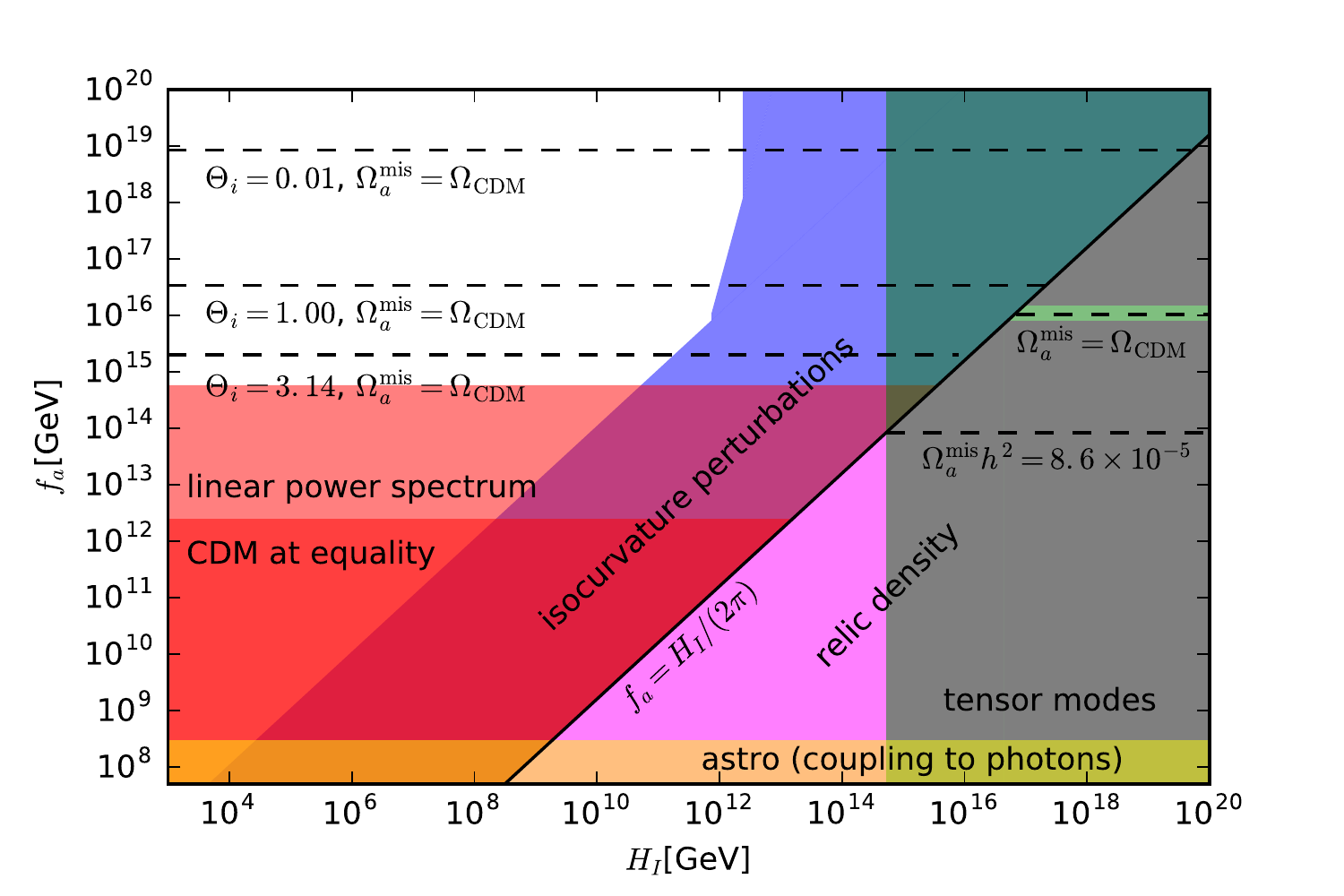}
 \caption{{\bf Cosmological Constraints on the Model.} {\it Left panel:} Constraints on the parameter space for the case in which the mass of the scalar particle does not change 
 with the temperature, $n=0$ in Eq.~(\ref{eq:mass}). {\it Right panel:} Same as in previous panel but for the case in which $n\gg 1$. In both cases we have fixed 
 the axion mass to $m_a=10^{-22}\,\textrm{eV}$. Note that current cosmological observations exclude Scenario~A, no matter the details of the mechanism giving rise to the axion mass.
 Scenario~B, on the contrary, is still viable as long as the initial value of the misalignment angle is chosen appropriately (white area on the top-left quadrants of the two panels).
 Note that the constraints from the linear power spectrum only affect fine tuned scenarios.}
 \label{fig:1}
\end{figure*}

\section{Cosmological and Astrophysical Constraints}\label{sec.constraints}

The main results of this paper are summarized in Figure~\ref{fig:1}, where we show, for the two limiting cases in Eq.~(\ref{eq:mass}), $n=0$ and $n\gg1$, the constraints that 
appear on the parameters of the model if an axion-like particle of mass $m_a\sim 10^{-22}\,\textrm{eV}$ constitutes the totality of the DM in the universe.

Note that the figures are divided by a transverse line, $f_a=H_I/(2\pi)$, in two different regions, corresponding to the two different analysis carried out in 
Sections~\ref{sec:after} and~\ref{sec:before}. We should look at these two regions in a different way. 

For the parameters below the line $f_a=H_I/(2\pi)$, Scenario~A, the PQ symmetry is broken after the end of inflation, and the quantity $\langle F_{n_a}\theta_i^2\rangle$ 
in Eq.~(\ref{eq:Omega}) is fixed to a constant value, see Eq.~(\ref{eq:scenarioA}) and the discussion below. In this case the amount of mass density in pNG bosons depends only 
on the axion decay constant, $f_a$, and not on the value of the Hubble parameter at the end of inflation, $H_I$. The upper dashed line 
to the right of the curve $f_a=H_I/(2\pi)$ corresponds to the particular value of the parameter $f_a$ such that the energy density in the misaligned vacuum 
accounts for the totality of the observed CDM in the universe, $\Omega_a^{\textrm{mis}}h^2=\Omega_{\textrm{CDM}}h^2\approx 0.12$~\cite{Planck}. For values of $f_a$ above this 
line we have $\Omega_a^{\textrm{mis}}>\Omega_{\textrm{CDM}}$, and then this region of the parameter space is excluded by observations.
For values of $f_a$ below the dashed line, on the contrary, we have $\Omega_a^{\textrm{mis}}<\Omega_{\textrm{CDM}}$, and then the energy density in the misaligned 
vacuum cannot account for the totality of the CDM. 

For the parameters above the line $f_a=H_I/(2\pi)$, Scenario~B, the pNG boson is formed before the end of inflation, and the value of 
$\langle F_{n_a}\theta_i^2 \rangle$ in Eq.~(\ref{eq:Omega}) is essentially an arbitrary positive definite constant, see Eq.~(\ref{eq:scenarioB}) above. 
Note that in this scenario the amount of mass density in axions depends not only on the scale energy $f_a$, but also on the value of the Hubble constant at the 
end of inflation, $H_I$. Now, for a particular value of the axion decay constant $f_a$, we can always fix the initial misalignment angle $\theta_i$ in the appropriate way such 
that the mass density in scalar particles of mass $m_a\sim 10^{-22}\,\textrm{eV}$ constitutes the totality of the CDM in the universe, $\Omega_a^{\textrm{mis}}h^2=\Omega_{\textrm{CDM}}h^2\approx 0.12$. 
Horizontal dashed lines to the left of the curve $f_a=H_I/(2\pi)$ represent particular realizations. 

The shaded areas in Figure~\ref{fig:1} represent regions in the parameter space that have been excluded by different observations.
In particular, the area colored in green comes from the amplitude of the (non-observed) tensor modes, $H_I<5.206\times 10^{14}\,\textrm{GeV}$, see Eq.~(\ref{eq:WMAP.HI}), 
whereas the area in blue from the amplitude of the (also non-observed) isocurvature perturbations, $H_I/(|\theta_i|f_a)<2.885\times 10^{-5}$, see Eq.~(\ref{eq:isocurvature}).
More details about inflation and its relation to the cosmological observables can be found in Appendix~\ref{appendix}.

We can also impose a very loose bound on the axion decay constant by demanding that, on linear scales, 
$k<0.1\,h\,\textrm{Mpc}^{-1}$, the mass power spectrum should not deviate significantly from the standard CDM prediction (far stronger constraints can be obtained in the nonlinear regime, 
but this requires more modeling uncertainty). Fixing the dimensionless density parameter in axions to 
$\Omega_a^{\textrm{mis}}h^2=\Omega_{\textrm{CDM}}h^2\approx 0.12$, we obtain $f_a\gtrsim 2.5m_{22}^{-1}\times 10^{14}\,\textrm{GeV}$ for the case of a $T$-independent mass, $n=0$, 
whereas $f_a\gtrsim 5.8m_{22}^{-1}\times 10^{14}\,\textrm{GeV}$ if $n\gg 1$. 

Note that this constraint is tighter than the one that appears by imposing a CDM background at 
matter-radiation equality, $T_{\textrm{osc}}\lesssim 0.5\,\textrm{eV}$, while keeping the condition of a total DM axion content, 
$f_a\gtrsim 1.4m_{22}^{-1/2}\times 10^{12}\,\textrm{GeV}$ if $n=0$, and $f_a\gtrsim 2.5m_{22}^{-1}\times 10^{12}\,\textrm{GeV}$ for a potential with a sharp $T$-dependence, $n\gg1$.
These constraints appear delimited in red in Figure~\ref{fig:1} with the labels ``linear power spectrum'' and ``CDM at equality'', respectively.

Compare this with the much lower astrophysical bounds arising from a possible coupling to photons, $g_{a\gamma}\lesssim 5\times 10^{-12}\,\textrm{GeV}$, as suggested by an analysis of the 
(lack of a) gamma ray signal from supernova SN1987a~\cite{Payez}; see Section~9 in Ref.~\cite{review} for more details on other nongravitational constraints. Taking $g_{a\gamma}=\alpha_{\textrm{EM}}/(2\pi f_a)$, 
where $\alpha_{\textrm{EM}}=1/137$ is the fine-structure constant, we obtain $f_a\gtrsim 2\times 10^8\,\textrm{GeV}$. This bound appears in yellow in Figure~\ref{fig:1}.

Finally, the area colored in magenta in the bottom-right quadrants of the two panels of Figure~\ref{fig:1} represent the parameter space that is excluded by relic density 
observations (over- or infra-production of axions).

If PQ symmetry breaking occurs after the end of inflation, $f_a<H_I/(2\pi)$, there is a region of the parameter space $(H_I,f_a)$ allowed after imposing constraints 
from tensor modes. Comparing the left and right panels of Figure~\ref{fig:1}, we see that this imposes $\Omega_ah^2\lesssim 9.9\times 10^{-7}$ in the $n=0$ case, and 
$\Omega_ah^2\lesssim 8.6\times 10^{-5}$ if $n\gg 1$. Thus, in Scenario A, the particles in the misaligned vacuum cannot provide a 
significant amount of the relic density, and this mechanism cannot explain the origin of the coherent state considered in the simulations of Refs.~\cite{simulations}.

Other contributions to the mass density in the form of string decays and/or thermal relics could also exist if the PQ symmetry is restored after the end of inflation. 
Ultralight axion thermal relics would contribute as hot DM, and they are not relevant for our purposes in this paper.
Remnants from cosmic defects can however contribute as CDM, but its abundance is usually expected to be of order one when compared to the misalignment production.
Increasing the mass of the particle to $m_a\sim 10^{-16}\,\textrm{eV}$ can make it possible to reach the critical density expected for a DM component in Scenario~A, but loosing the success 
of the model when addressing the small scale discrepancies of the standard CDM paradigm.

On the other hand, if PQ symmetry is broken during inflation, $f_a>H_I/(2\pi)$, then an ultralight axion of mass $m_a\sim 10^{-22}\,\textrm{eV}$ 
can constitute the totality of the observed DM in the universe, as long as the initial misalignment angle is chosen properly
and inflation occurs at a scale $H_I\lesssim 2.5\times 10^{12}\,$GeV. 
Such a low value of the Hubble parameter corresponds to a tensor-to-scalar ratio of order $r\lesssim 10^{-5}$, inaccessible to near-future ground-based 
CMB telescopes~\cite{s4scibook}.\footnote{Delensing the CMB can lead to substantial improvements on constraints to the scalar-to-tensor ratio~\cite{s4scibook} and other cosmological 
parameters~\cite{delensing}, and combined with the (to be observed) 21cm power spectrum could, in principle, give access to values of this ratio as low as $r\sim 10^{-9}$~\cite{21cm}.} 
This conclusion is not sensitive to the details of the phase transition at the scale of the explicit symmetry breaking,
as can be appreciated in the two limiting cases illustrated in the two panels of Figure~\ref{fig:1}. Notice 
that if the axion mass does not depend on the temperature, and in order to avoid a fine-tuned initial condition around $\theta_i=\pi$,
we need $f_a$ to lie not far from the Planck scale, $f_a\gtrsim 5\times 10^{-3}m_{\textrm{Pl}}$. In terms of the the explicit symmetry breaking that implies $1\,\textrm{eV}<\mu<100\,\textrm{eV} $,
around the scale of neutrino physics. This is an order of magnitude lower than the value expected from the harmonic approximation. 
Temperature dependent effects can easily decrease this lower value even another order of magnitude, see right panel in Figure~\ref{fig:1}.

However, in general, an instanton potential and a $T$-dependent axion mass delay the onset of field oscillations, and this can only increase a mild tension previously
identified in Ref.~\cite{Alma} from the comparison of the results obtained by analyzing independently observations carried out at different scales.
On the one hand, high redshift galaxies suggest a lower bound for the axion mass of $m_a^{\textrm{cos}}\gtrsim 1\times 10^{-22}\,\textrm{eV}$~\cite{high_z} (or even higher~\cite{high_z_new}), whereas a detailed analysis of the internal stellar dynamics in dwarf 
spheroidals find a better agreement with $m_a^{\textrm{astro}}<0.4\times 10^{-22}\,\textrm{eV}$~\cite{Alma}.
The possible nonlinear, $T$-dependent contributions associated to a realistic axion potential can only increase the value of $m_a^{\textrm{cos}}$ while leaving 
$m_a^{\textrm{astro}}$ unaffected, aggravating the situation even more. 
This seems to suggest that, in practice, any nonlinear and/or temperature-dependent imprint of the potential
term into the cosmological observables should be very small, at least if the core/cusp and missing satellites problems want to be addressed simultaneously.

For the case of a string axion, or an accidental one, the potential appears naturally $T$-independent, and there should not be any 
signal unless the initial misalignment angle were unnaturally fine tuned around the critical value $\theta_i= \pi$, and then $f_a\lesssim 10^{16}\,\textrm{GeV}$.
In the case of a mass term generated through the transition to a strong coupling regime, however, a $T$-dependent potential could have leaved some interesting
imprints on larger $f_a$ scenarios, accessible only through observations of the nonlinear mass power spectrum.

\section{Discussion and Conclusions}\label{sec.discussion}

The fuzzy DM model is today both popular and intriguing. Quantum mechanics together with current cosmological observations point to $m\gtrsim 10^{-22}\text{ eV}$ as an absolute lower bound for 
the mass of a DM particle. Exploring the edge of this bound offers the possibility to observe exotic wave-like effects on galaxy formation, which could one day lead to a preference for 
fuzzy over CDM. Halo density profiles are only sensitive to the zero-temperature mass of an axion-like particle. However, possible interactions with standard model particles, 
and with the axion field itself, are determined by a second parameter, the axion decay constant, $f_a$. In this paper we have determined the range of possible interaction strengths for a fuzzy 
DM component using an instanton model for the axion potential and a possible temperature dependence of the mass, applying constraints from relic density, structure formation, 
and CMB tensor and isocurvature modes.

An ultralight axion could be realized in a natural manner if a global PQ-like symmetry is ``accidental'', descending from a fundamental discrete 
symmetry~\cite{KimMarsh}, or localized in a hidden sector version of QCD with a low confinement scale~\cite{ULA_newmodel}. In the first of these cases the axion mass 
is effectively $T$-independent, while in the second one non-perturbative effects ``switch on'' the mass at some low energy scale $\mu$, with $m_a\propto (T/\mu)^{-n}$
at temperatures of order $T\sim \mu$.

For an axion in field theory we expect the global PQ symmetry to be linearly realized at temperatures above $f_a$, allowing for the possibility of a late-time phase transition 
such as that described in Scenario A. However, for ultralight axions, this scenario is excluded by relic density and tensor mode constraints, and so we expect to see no phase transition 
remnants such as ``axion miniclusters'' from large amplitude perturbations imprinted by the breakdown of the symmetry~\cite{miniclusters}, and no population of cold axions 
from string decays~\cite{string_decay}. It is worth noting, however, that these conclusions can be substantially altered in non-minimal multi-field~\cite{nflation} and ``clockwork axion'' 
models~\cite{clockwork}, and possibly also in the context of axion monodromy DM~\cite{monodromy}.

If the symmetry is broken during inflation, Scenario B, and also in the light of a field theory, the axion particles can instead contribute to conform the totality of the CDM,
provided the initial value of the misalignment angle is chosen appropriately. If in addition the variation of the axion mass with the temperature is sharp around the scale
of the breakdown of the shift symmetry, $T\sim \mu$, we find that $f_a>10^{15}\,\textrm{GeV}$ neglecting a severe fine tuning initial condition; see the top-left quadrant of the right panel
of Figure~\ref{fig:1}.

Neither late time symmetry breaking of Scenario~A, nor a $T$-dependent axion mass are expected for the abundant closed-string axions of string theory. The PQ symmetry is 
never linearly realized in 3+1-dimensions, and non-perturbative effects generically switch on at a UV scale $\Lambda_{\rm UV}\gg \mu$~\cite{axiverse}. We thus expect ultralight 
string axions to live in the top-left quadrant of the left panel of Figure~\ref{fig:1}, requiring larger values of $f_a>10^{16}\text{ GeV}$, which are otherwise obtained naturally in 
canonical small volume compactifications~\cite{small-volume, Acharya}. 
  
Axion couplings to ordinary matter scale as $g\propto 1/f_a$, and so the lower values of $f_a\sim 10^{15}\,\text{ GeV}$ allowed by current cosmological observations
would simplify direct detection (by e.g. CASPEr-Wind-like experiments~\cite{casper,KimMarsh}) when compared to previous estimates obtained under the harmonic potential 
approximation. The later onset of axion oscillations in low $f_a$ scenarios could also lead to interesting effects in structure formation on scales accessible to CMB 
lensing~\cite{Hlozek16}, but probably at the expense of some desirable features in galaxy formation arising from the well studied zero-temperature potential. 
Isocurvature perturbations could be detected alongside effects on structure formation if $H_I\sim 10^{12}\,\text{GeV}$, while an observation 
of tensor modes in the near future would seem to rule out the possibility of an ultralight axion DM component~\cite{s4scibook,Marshbicep}. A coupling of the axion to photons could also induce small-angle 
$B$-mode polarization via the birefringent effect, which dominates over lensing $B$-modes at small angular scales~\cite{Liu}

The axion decay constant $f_a$ also sets the strength of axion quartic self-interactions, which affects the stability of nonlinear compact configurations known as 
``axion stars.'' Axion stars of masses above the critical value
\begin{eqnarray}\label{eq.crit.stars}
M_{\rm cr, stars} &=& 10.2\, \frac{f_a M_{\textrm{Pl}}}{m_a} \\
&=&  1.1\times 10^{10}M_\odot\left(\frac{f_a}{10^{16}\,\text{GeV}} \right)\left(\frac{10^{-22}\,\text{eV}}{m_a}\right) \nonumber \label{eqn:critical}
\end{eqnarray}
have been found to be unstable, ejecting a significant amount of their super-critical mass as relativistic axions~\cite{Levkov,Helfer}. Such a process would convert a fraction 
of the total cold axion density generated through vacuum realignment into hot or warm DM. If super-critical axion stars can form in abundance, then this process could offer 
new constraints or observational signatures of the model. 

In order to illustrate this possibility, let us consider the critical mass when applied to the axion star cores of DM halos. At redshift zero, numerical cosmological simulations 
point to a correlation between the core- and halo-masses in galaxies of the form~\cite{simulations} 
\be
M_{\rm core} = 1.1\times 10^7 M_\odot \left(\frac{M_{\rm halo}}{4.4\times 10^7 M_\odot}\right)^{1/3} \left(\frac{m_a}{10^{-22}\,\text{eV}}\right)^{-3/2} .
\ee
Extrapolating this relation up to the critical mass in Eq.~(\ref{eq.crit.stars}) suggests that DM halos above
\be
M_{\rm cr,halo}=4.4\times 10^{16}M_\odot \left(\frac{f_a}{10^{16}\,\text{GeV}} \right)^3\left(\frac{m_a}{10^{-22}\,\text{eV}}\right)^{3/2} 
\ee
would contain unstable cores. The most massive clusters observed have masses of order $M_{\rm halo}\sim 3\times 10^{15}M_{\odot}$~\cite{Menanteau}, suggesting that with 
$m_a\sim 10^{-22}\,\text{eV}$ and $f_a\sim 10^{16}\,\text{GeV}$ the axion star instability does not affect any known DM halos. However, with $f_a\sim 10^{15}\,\text{GeV}$, 
galaxy clusters not much more massive than our own Milky Way will contain unstable axion star cores, as pointed out in Ref.~\cite{Levkov}.  

To make $f_a$ as low as $10^{15}\,\text{GeV}$ a combination of $n\gg 1$ and a tuning of the initial misalignment to 
$\theta_i-\pi\sim 10^{-3}$ is required, see the top-left quadrant of the right panel of Figure~\ref{fig:1}, thus  
suggesting that axion star instability in DM halo cores is on the edge of the allowed parameter space for $m_a\sim 10^{-22}\text{ eV}$.

Another possibility to excite macroscopically the zero mode of a scalar particle, in this case thermally, is through the formation
of a Bose-Einstein condensate in a universe with an asymmetry between the number densities of scalar particles and antiparticles~\cite{BEC}.
We discuss this scenario in more detail in Ref.~\cite{BEC2}.

\acknowledgments

We are grateful to Patrick Draper and Ren\'ee Hlozek for useful comments and discussions, and also to Anthony Aguirre for collaborating during the first stages of this project.
ADT is supported in part by CONACyT Mexico under Grants No.~182445 and No.~167335, and Grant No.~FQXi-1301 from the Foundational Questions
Institute (FQXi). DJEM acknowledges support of a Royal Astronomical Society Postdoctoral Fellowship hosted by King's College London.

\appendix

\section{Anharmonic Corrections to the Relic Density and Onset of Field Oscillations}\label{appendix2}

\begin{figure*}
\begin{center}
$\begin{array}{@{\hspace{-0.1in}}l@{\hspace{-0.1in}}l}
\includegraphics[width=\columnwidth]{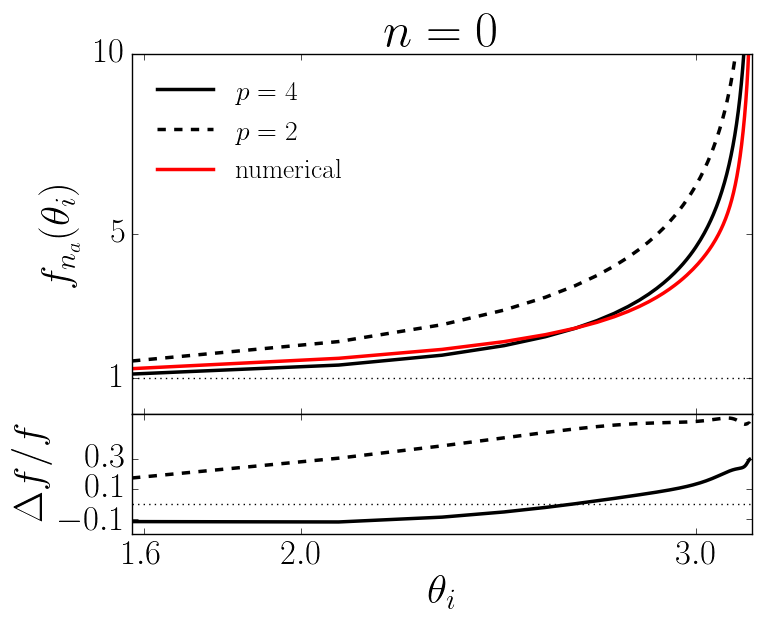} \hspace{0.8cm}
\includegraphics[width=\columnwidth]{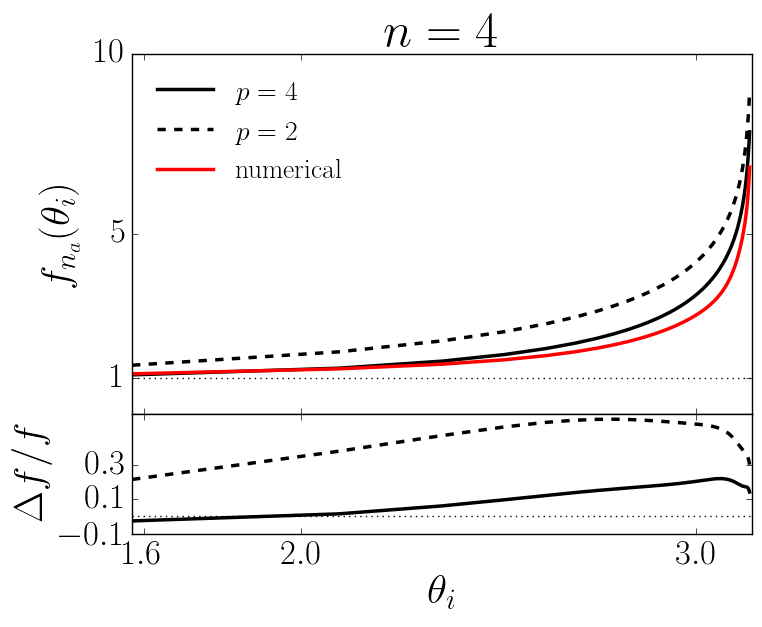}
 \end{array}$
 \end{center}
\caption{{\bf Anharmonic Corrections to the Relic Density.} 
{\it Left panel:} The anharmonic correction function $f_{n_a}(\theta_i)$ computed from numerical solutions of Eqs.~(\ref{eq.KGanhar}) and~(\ref{eq.KGhar}), red line,
versus a fitting formula parameterized by the polynomial $q(x)=1-x^p$ in Eq.~(\ref{eqn:anharmonic_time}), for the case in which $n=0$. 
{\it Right panel:} Same as in previous panel but for the case in which $n=4$. Note that $p=4$ (continuous line) provides a better fit than $p=2$ (dotted line) to the red curve in both cases.
}
\label{fig:anharmonic}
\end{figure*}

We compute the anharmonic corrections by solving the equation of motion
\be\label{eq.KGanhar}
\ddot{\theta}+3\tilde{H}(\tilde{t})\dot{\theta}+\tilde{m}_a^2(\tilde{t})\sin \theta = 0, 
\ee
and comparing the results to those obtained under the harmonic approximation, 
\be\label{eq.KGhar}
\ddot{\theta}+3\tilde{H}(\tilde{t})\dot{\theta}+\tilde{m}_a^2(\tilde{t})\theta = 0. 
\ee
Here $\tilde{t}$, $\tilde{H}$, and $\tilde{m}_a$ are dimensionless quantities measured in units of $m_a$, the mass of the particle at zero-temperature, and the dot denotes the 
derivative with respect to the normalized time. From now on and for convenience we will drop the tildes in the expressions below. 
Since for the masses of interest $t_{\rm osc}$ is always sufficiently large in physical units, we can treat the number of relativistic degrees of freedom 
as constant in Eqs.~(\ref{eq.KGanhar}) and~(\ref{eq.KGhar}). 
The temperature is then proportional to the inverse of the scale factor, $T\propto 1/a$. During radiation domination $a\propto 1/t^{1/2}$, with the Hubble rate and the axion mass scaling in time like $H\propto 1/t$
and $m_a \propto t^{n/2}$, respectively. 

The effect of the anharmonic corrections can be shown to lead an increasing time for the onset of field oscillations that is logarithmically divergent~\cite{Lyth}:
\be
 m_a(t_{\rm osc}) t_{\rm osc} = \ln [e/q(x)] \, ,
\label{eqn:anharmonic_time}
\ee
where $q(x)$ is a polynomial function of $x=\theta_i/\pi$, see below for details. The use of a polynomial argument instead of the original result derived by Lyth in Ref.~\cite{Lyth} gives rise to the 
appropriate limiting behavior. 

We numerically solve the anharmonic and harmonic equations up to $t_f=10t_{\rm osc}$, well inside the adiabatic regime, such that the energy density, 
$\rho=\frac{1}{2}f_a^2\dot{\theta}^2+V(f_a\theta)$, in both cases scales as $\rho\propto a^{-3}$ and the ratio $f_{n_a}(\theta_i)=\rho_{\textrm{anh}}/\rho_{\rm har}$ remains 
constant (this is essentially the function $F_{n_a}$ in Eq.~(\ref{eq.deffna}) with the $T$-dependent term factorized out). This ratio is then evaluated for a large number of values of $\theta_i$. The results are shown in Figure~\ref{fig:anharmonic} for evolutions of the axion mass with 
temperature $n=0$ and $n=4$.
Ref.~\cite{Gondolo1} takes $q(x)=1-x^2$. In Figure~\ref{fig:anharmonic} we show the fitting function $q(x)=1-x^p$ for $p=2$ and $p=4$;
this point to a better fit to our numerical results taking $q(x)=1-x^4$. We use this choice in the main body of the text to compute the relic density, Eq.~(\ref{eq.def.f}). 

\section{Inflation and the Cosmological Observables}\label{appendix}

The power spectrum of curvature, $\Delta^2_{\mathcal{R}}(k)$, and tensor, $\Delta^2_{h}(k)$, perturbations are related to the Hubble rate at the end of inflation, $H_I$,
through the expressions
\begin{equation}
\label{eq:A.power}
 \Delta^2_{\mathcal{R}}(k)=\frac{H_I^2}{8\pi^2M_{\textrm{Pl}}^2\epsilon} , \quad \Delta^2_{h}(k)=\frac{2 H_I^2}{\pi^2M_{\textrm{Pl}}^2}.
\end{equation} 
Here $M_{\textrm{Pl}}$ is the Planck mass and $\epsilon$ a slow-roll parameter. (There is a small dependence of 
the power spectrum of curvature and tensor perturbations on the wavenumber $k$, however, this variation is not going to be relevant for our purposes in this paper.)
The \emph{Planck} collaboration~\cite{Planck} has reported curvature perturbations at fixed wavenumber $k_*=0.05\,\textrm{Mpc}^{-1}$ of amplitude
\begin{equation}
\label{eq:A.power.R.WMAP}
 \Delta^2_{\mathcal{R}}(k_*)=(2.196\substack{+0.080 \\ -0.076})\times 10^{-9}
\end{equation}
at $68\%$ confidence level (CL). This quantity alone does not determine the scale energy of inflation, $H_I$, since we do not have access to the 
value of the slow-roll parameter $\epsilon$. However, \emph{Planck} reports also the absence of tensor perturbations with amplitude higher than
\begin{equation}
\label{eq:A.r}
 r(k_*)\equiv\frac{\Delta^2_{h}(k_*)}{\Delta^2_{\mathcal{R}}(k_*)}<0.168
\end{equation}
at $95\%$CL. Combining Eqs.~(\ref{eq:A.power}), (\ref{eq:A.power.R.WMAP}), and~(\ref{eq:A.r}) we obtain an upper limit for the scale energy of inflation,
\begin{equation}
\label{eq:WMAP.HI}
 H_I<5.206\times10^{14}\,\textrm{GeV}.
\end{equation}

In addition to curvature perturbations, the presence of light bosons during the epoch of inflation would leave isocurvature fluctuations on the power spectrum 
of primordial perturbations. For the case of a massless particle, like it would be the case for any pNG boson at the scale energy $H_I$, we obtain~\cite{Gondolo1}
\begin{equation}
\label{eq:A.isocurvature.power}
 \Delta_a^2=\frac{H_I^2}{\pi^2\theta_i^2f_a^2}.
\end{equation}
However, isocurvature perturbations are tightly constrained by the observations. For instance, \emph{Planck} reports the absence of isocurvature perturbations 
with an adiabaticity factor
\begin{equation}
\label{eq:A.adiabaticity}
 \frac{\Delta^2_{a}(k_*)}{\Delta^2_{\mathcal{R}}(k_*)}\equiv\frac{\alpha_0(k_*)}{1-\alpha_0(k_*)}
\end{equation}
higher than 
\begin{equation}
\label{eq:A.adiabaticity.WMAP}
 \alpha_0(k_*)< 0.037 
\end{equation}
at $95\%$CL. Combining Eqs.~(\ref{eq:A.power.R.WMAP}), (\ref{eq:A.isocurvature.power}), (\ref{eq:A.adiabaticity}), and~(\ref{eq:A.adiabaticity.WMAP}) we obtain the upper limit
\begin{equation}
\label{eq:isocurvature}
 \frac{H_I}{|\theta_i| f_a}<2.885\times 10^{-5},
\end{equation}
for the ratio of the scale energy of inflation to the scale energy of the spontaneous symmetry breaking multiplied by the initial misalignment angle.

[Note that similar values are also obtained using the results by the WMAP9~\cite{WMAP9} collaboration combined with eCMB+BAO+$H_0$:
$\Delta^2_{\mathcal{R}}(k_0)=(2.464\pm0.072)\times 10^{-9}$, $r(k_0)<0.13$, and $\alpha_0(k_0)<0.047$, where now  $k_0=0.002\,\textrm{Mpc}^{-1}$
and using the same interval of CL.]

\end{document}